\documentclass[conference]{IEEEtran}
\usepackage[T1]{fontenc}
\usepackage[latin9]{inputenc}
\usepackage[table]{xcolor}
\usepackage{colortbl}
\usepackage{amsmath}
\usepackage{amssymb}
\usepackage{stackrel}
\usepackage{graphicx}

\makeatletter

\providecommand{\tabularnewline}{\\}

\usepackage[caption=false,font=footnotesize]{subfig}

\makeatother

\begin{document}
\title{Fast Antenna Coding Based on Pixel Antennas for Enhancing Spectral
Efficiency of MIMO Systems}
\author{\IEEEauthorblockN{Zixiang Han\IEEEauthorrefmark{1}, Xin Su\IEEEauthorrefmark{1}, Jing
Jin\IEEEauthorrefmark{1}, Shanpu~Shen\IEEEauthorrefmark{2}, Qingqing
Wu\IEEEauthorrefmark{3}, Yifei Yuan\IEEEauthorrefmark{1}, Jiangzhou
Wang\IEEEauthorrefmark{4}, and Chih-Lin I\IEEEauthorrefmark{1}}\IEEEauthorblockA{\IEEEauthorrefmark{1}Future Research Lab, China Mobile Research Institute,
100053, Beijing, China\\
\IEEEauthorrefmark{2}State Key Laboratory of Internet of Things for
Smart City, University of Macau, Macau, China\\
\IEEEauthorrefmark{3}Department of Electronic Engineering, Shanghai
Jiao Tong University, Shanghai, China\\
\IEEEauthorrefmark{4}School of Information Sciences and Engineering,
Southeast University, Nanjing, China.\\
Email: hanzixiang@chinamobile.com}}
\maketitle
\begin{abstract}
A fast antenna coding technique based on pixel antennas is investigated.
Antenna coding is a technique that dynamically controls the ON/OFF
states of PIN diodes to adjust radiation patterns of pixel antennas.
By switching the antenna coder at the baseband symbol period, this
technique can modulate information on the radiation patterns in the
beamspace. This allows the pixel antenna to transmit additional information,
thereby enhancing the spectral efficiency (SE) of multiple-input multiple-output
(MIMO) systems. Specifically in this work, we firstly introduce the
concept of pixel antenna and the fast antenna coding. Then based on
the beamspace channel representation, we provide the system model
of MIMO with the fast antenna coding and derive the SE and bit error
rate (BER). The SE enhancement achieved by fast antenna coding is
also demonstrated. An optimization approach is also presented for
obtaining the antenna coders that can generate the desired radiation
patterns for fast antenna coding. Using the optimized antenna coders,
we evaluate the SE, energy efficiency (EE), and BER of MIMO systems
with fast antenna coding. The simulation results show that for a 4-by-4
MIMO system, SE is enhanced by 12 bits/s/Hz and EE can be improved
by 33\% when compared with conventional MIMO, demonstrating the effectiveness
of the fast antenna coding technique.
\end{abstract}


\section{Introduction}

Multiple-input multiple-output (MIMO) has greatly enhanced the spectral
efficiency (SE) of wireless communication by leveraging the spatial
degrees of freedom (DoF) \cite{1353475}. Antennas of conventional
MIMO have fixed geometries and radiation patterns so that enhancing
SE of conventional MIMO relies on stacking more antennas. However,
the cost and energy consumption of MIMO systems have approached the
limits in practical implementations, resulting in a bounded SE. Therefore,
to serve the growing data traffic, the upcoming 6G mobile network
demands novel antenna techniques to provide extra DoF for breaking
through the SE limits.

One promising technology to address this issue is pixel antennas which
are composed by a set of sub-wavelength elements, denoted as pixels,
and a group of radio frequency (RF) switches embedded across each
pair of adjacent pixels \cite{1367557}. By controlling switch states,
the pixel antenna topology is reconfigured and the resulting radiation
patterns can be adjusted. A variety of pixel antennas have been designed
to implement pattern reconfigurability, which can steer the main beam
without the usage of phase shifters \cite{Zhang2022a}, \cite{Zhang2022}.
However, these designs aim to enhance the antenna performance from
hardware design perspective while the capability of pixel antennas
in enhancing spatial multiplexing of MIMO systems is not investigated.

The fluid antenna system (FAS) is an emerging technique which aims
to create a favorable propagation channel via changing antenna position
so that the multiplexing, diversity as well as beamforming gain can
be enhanced by FAS \cite{Wong2021}. FAS can be implemented based
on the mechanical movement or the fluidity of liquid material. Alternatively,
pixel antennas can mimic antenna position changes in FAS by controlling
switch states \cite{Wong2026}, which have preliminarily shown the
great potential for enhancing wireless communications.

Recently, a novel technique known as \textit{antenna coding} has been
proposed \cite{shen2024antenna}-\nocite{11494037}\nocite{11589539}\cite{Han2026},
where the states of RF switches in pixel antennas are characterized
by binary variables called the antenna coder which is optimized to
enhance the wireless system performance. Antenna coding can be categorized
into slow antenna coding and fast antenna coding. In the slow antenna
coding, antenna coders are switched according to the channel state
information so that radiation patterns of pixel antennas are adjusted
to improve the channel gain or suppress inter-user interference \cite{shen2024antenna},
\cite{11494037}. It can also be utilized to enhance the wireless
power transfer systems \cite{11589539}. On the other hand, in the
fast antenna coding, the switching interval of antenna coders is at
the symbol level, so that additional information can be mapped onto
the reconfigurable radiation patterns of pixel antennas for enhancing
SE of MIMO systems \cite{Han2026}. However, the previous work \cite{Han2026}
only considers switching antenna coders to excite orthogonal radiation
patterns, which limits the SE enhancement through fast antenna coding.

To overcome this limitation, in this work we further extend the antenna
coding design to a more general case beyond orthogonal radiation patterns
for the fast antenna coding, so that a higher SE can be achieved.
Specifically, we firstly introduce the concept of the pixel antenna
and antenna coding. Then we derive the SE and bit error rate (BER)
expressions of MIMO systems with the fast antenna coding. The SE enhancement
achieved by fast antenna coding is also demonstrated. An optimization
approach is also described for obtaining the desired antenna coders.
Utilizing the optimized antenna coders, SE, energy efficiency (EE),
and BER performance of MIMO systems implemented by pixel antennas
are provided to demonstrate the effectiveness of fast antenna coding.

\section{Antenna Coding Based on Pixel Antennas}

\begin{figure}[t]
\begin{centering}
\textsf{\includegraphics[width=8.5cm]{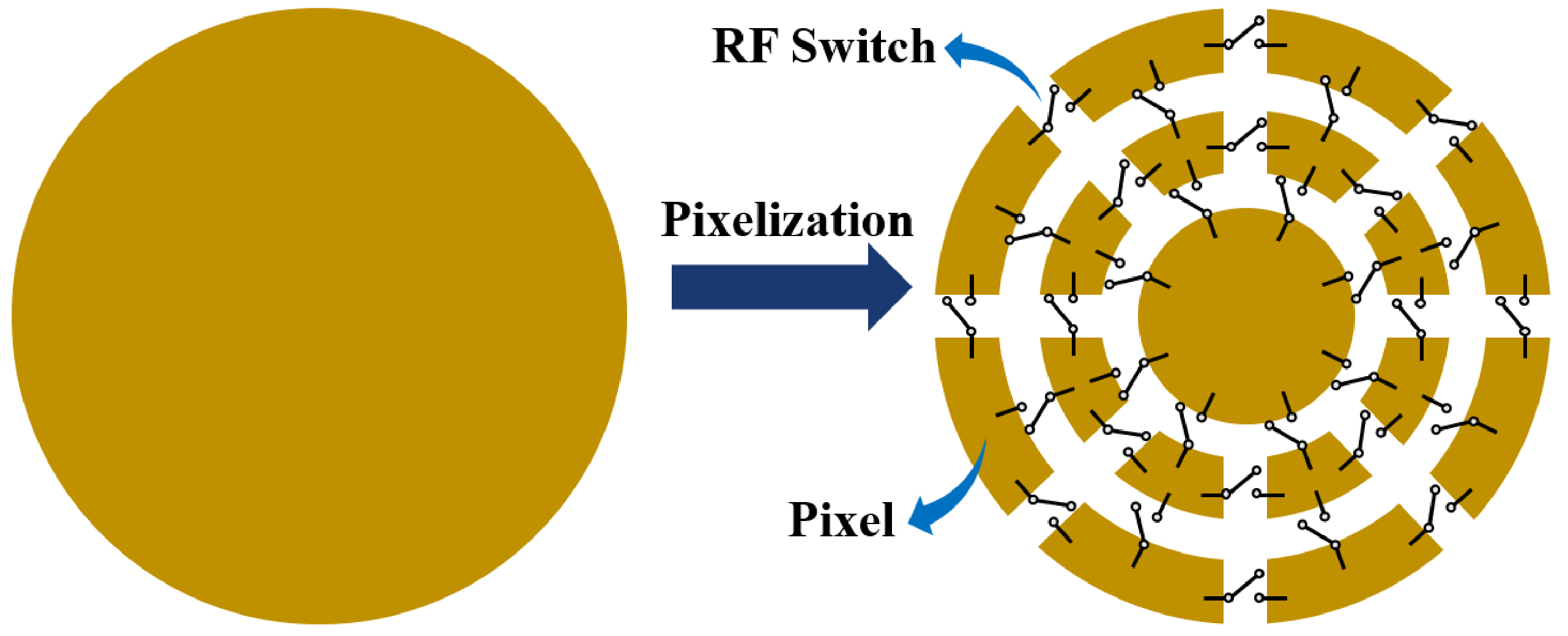}}
\par\end{centering}
\begin{centering}
(a)
\par\end{centering}
\begin{centering}
\textsf{\includegraphics[width=8.5cm]{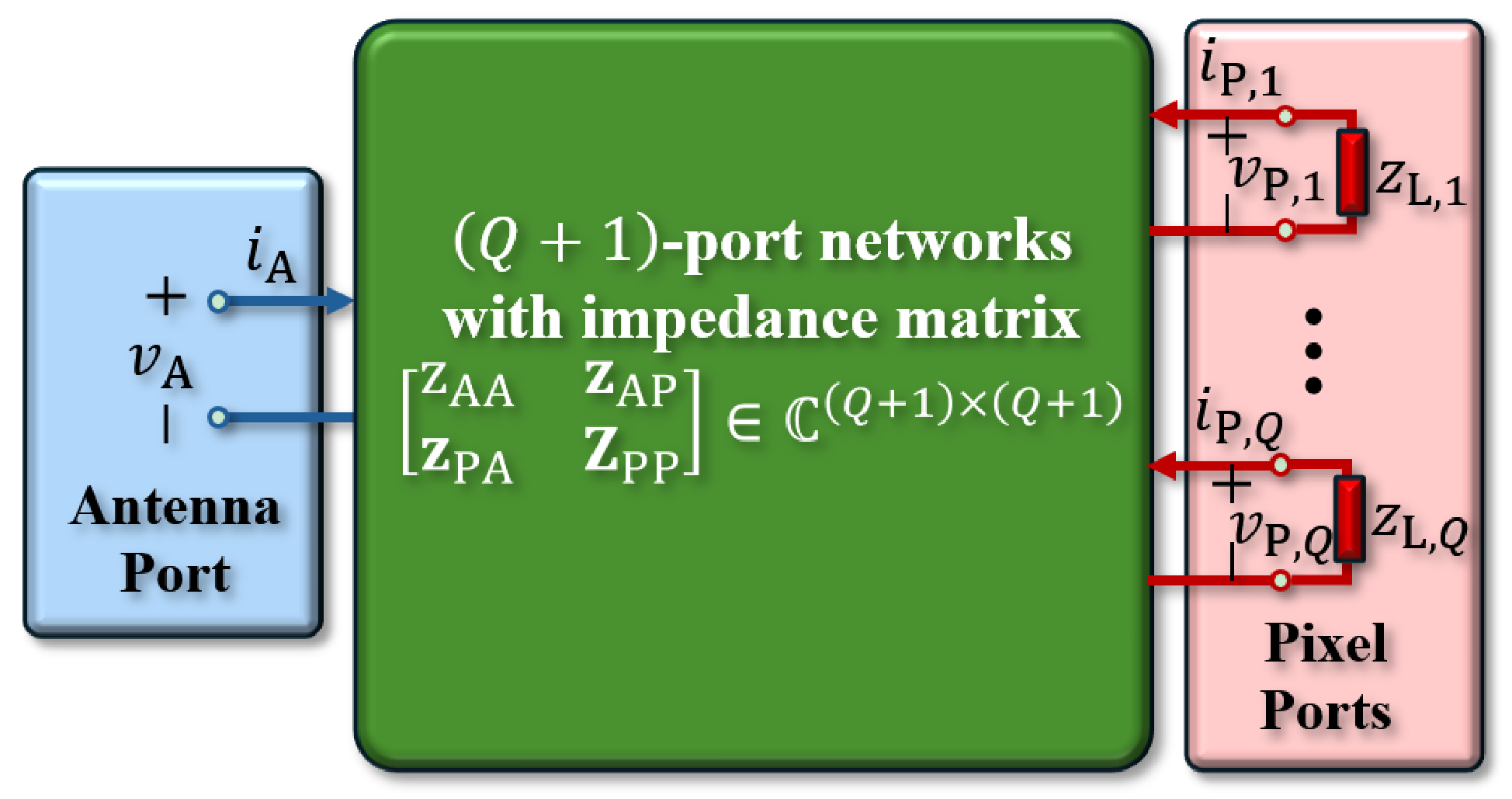}}
\par\end{centering}
\begin{centering}
(b)
\par\end{centering}
\caption{Schematic of (a) a pixel antenna and (b) the multiport circuit model.\label{fig:Array}}
\end{figure}
Pixel antenna is a general paradigm of reconfigurable antennas, which
can be viewed as discretizing a continuous radiating surface into
a grid of pixels and using variable lumped elements to connect adjacent
pixels. The shape of pixels can be arbitrary, rather than being limited
to squares. For example, as shown in Fig. \ref{fig:Array}(a), we
can discretize a circular radiator into a center circle and multiple
arc-shaped pixels. Adjacent pixels can be interconnected by RF switches.
A pixel antenna embedded with $Q$ switches can be formulated by a
$\left(Q+1\right)$-port network, as shown in Fig. \ref{fig:Array}(b),
where $Q$ pixel ports are placed across adjacent pixels for implementing
RF switches and a single antenna port is connected to an RF chain
for exciting the pixel antenna. The impedance matrix of the $\left(Q+1\right)$-port
network is written as
\begin{equation}
\mathbf{Z}=\left[\begin{array}{ll}
z_{\mathrm{AA}} & \mathbf{z}_{\mathrm{AP}}\\
\mathbf{z}_{\mathrm{PA}} & \mathbf{Z}_{\mathrm{PP}}
\end{array}\right]\in\mathbb{C}^{\left(Q+1\right)\times\left(Q+1\right)},\label{eq:Z}
\end{equation}
where $z_{\mathrm{AA}}\in\mathbb{C}$ refers to the impedance of the
single antenna port, $\mathbf{Z}_{\mathrm{PP}}\in\mathbb{C}^{Q\times Q}$
refers to the impedance matrix of the $Q$ pixel ports, $\mathbf{z}_{\mathrm{AP}}\in\mathbb{C}^{1\times Q}$
refers to the mutual impedance between the antenna port and pixel
ports, and $\mathbf{z}_{\mathrm{PA}}=\mathbf{z}^{T}_{\mathrm{AP}}$.
The states of $Q$ switches can be collected into a binary vector
$\mathbf{b}=\left[b_{1},b_{2},...,b_{Q}\right]^{T}\in\mathbb{\mathbb{R}}^{Q\times1}$,
$b_{q}\in\left\{ 0,1\right\} $, $\forall q$, which is referred to
as the antenna coder. Thus, the load impedance of $Q$ switches can
be written as a diagonal matrix $\mathbf{Z}_{\mathrm{L}}\left(\mathbf{b}\right)=\mathrm{diag}\left(z_{\mathrm{L},1},z_{\mathrm{L},2},...,z_{\mathrm{L},Q}\right)$,
where the load impedance $z_{\mathrm{L},q}$ models the impedance
of ON and OFF states for the $q$th RF switch as
\begin{equation}
z_{\mathrm{L},q}=\begin{cases}
0, & b_{q}=0,\:\textrm{i.e. switch\:ON}.\\
\infty, & b_{q}=1,\:\textrm{i.e. switch\:OFF}.
\end{cases}\label{eq:Switch}
\end{equation}

Exciting the antenna port by the current $i_{\mathrm{A}}\in\mathbb{C}$,
the current through the $Q$ pixel ports can be obtained by analyzing
the $\left(Q+1\right)$-port network \cite{pozar2011microwave}, written
as
\begin{equation}
\mathbf{i}_{\mathrm{P}}\left(\mathbf{b}\right)=-\left(\mathbf{Z}_{\mathrm{PP}}+\mathbf{Z}_{\mathrm{L}}\left(\mathbf{b}\right)\right)^{-1}\mathbf{z}_{\mathrm{PA}}i_{\mathrm{A}}.\label{eq:current relation}
\end{equation}
Hence, the radiation pattern of the pixel antenna is given by
\begin{align}
\mathbf{e}\left(\mathbf{b}\right) & =\mathbf{e}_{\mathrm{A}}i_{\mathrm{A}}+\mathbf{E}_{\mathrm{P}}\mathbf{i}_{\mathrm{P}}\left(\mathbf{b}\right)=\mathbf{E}_{\mathrm{oc}}\mathbf{i}\left(\mathbf{b}\right),\label{eq:Pattern}
\end{align}
where $\mathbf{E}_{\mathrm{oc}}=\left[\mathbf{e}_{\mathrm{A}},\mathbf{E}_{\mathrm{P}}\right]$
with $\mathbf{e}_{\mathrm{A}}\in\mathbb{C}^{2K\times1}$ and $\mathbf{E}_{\mathrm{P}}=\left[\mathbf{e}_{\mathrm{P},1},\ldots,\mathbf{e}_{\mathrm{P},Q}\right]\in\mathbb{C}^{2K\times Q}$
referring to the open-circuit radiation patterns of the antenna port
and pixel ports sampled at $K$ spatial angles for two polarization
components, and $\mathbf{i}\left(\mathbf{b}\right)=\left[i_{\mathrm{A}},\mathbf{i}^{T}_{\mathrm{P}}\left(\mathbf{b}\right)\right]^{T}$.
It can be observed that both current and radiation pattern in \eqref{eq:current relation}
and \eqref{eq:Pattern} can be coded by $\mathbf{b}$.

\begin{figure}[t]
\begin{centering}
\textsf{\includegraphics[width=8cm]{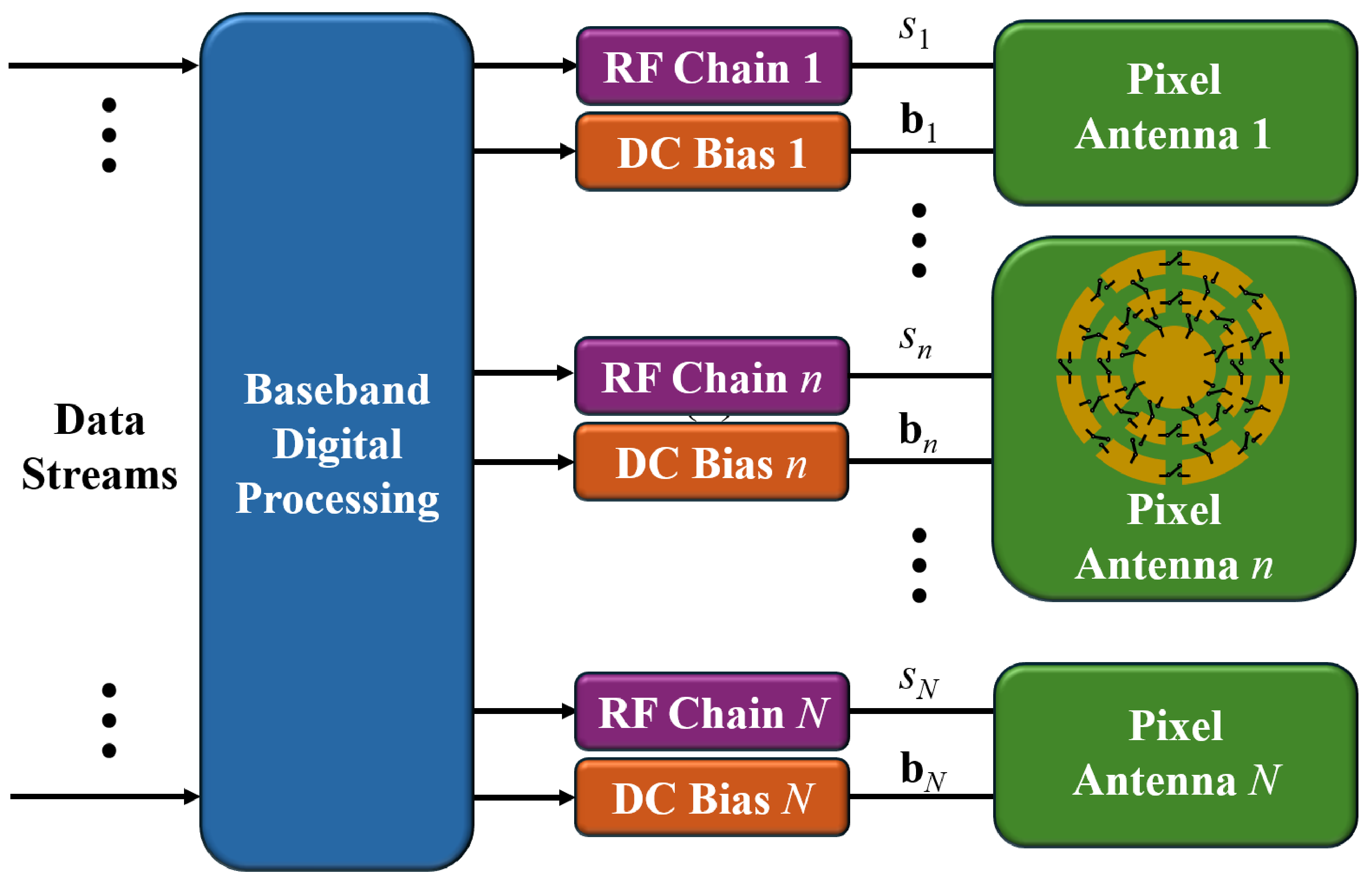}}
\par\end{centering}
\begin{centering}
(a)
\par\end{centering}
\begin{centering}
\textsf{\includegraphics[width=8.8cm]{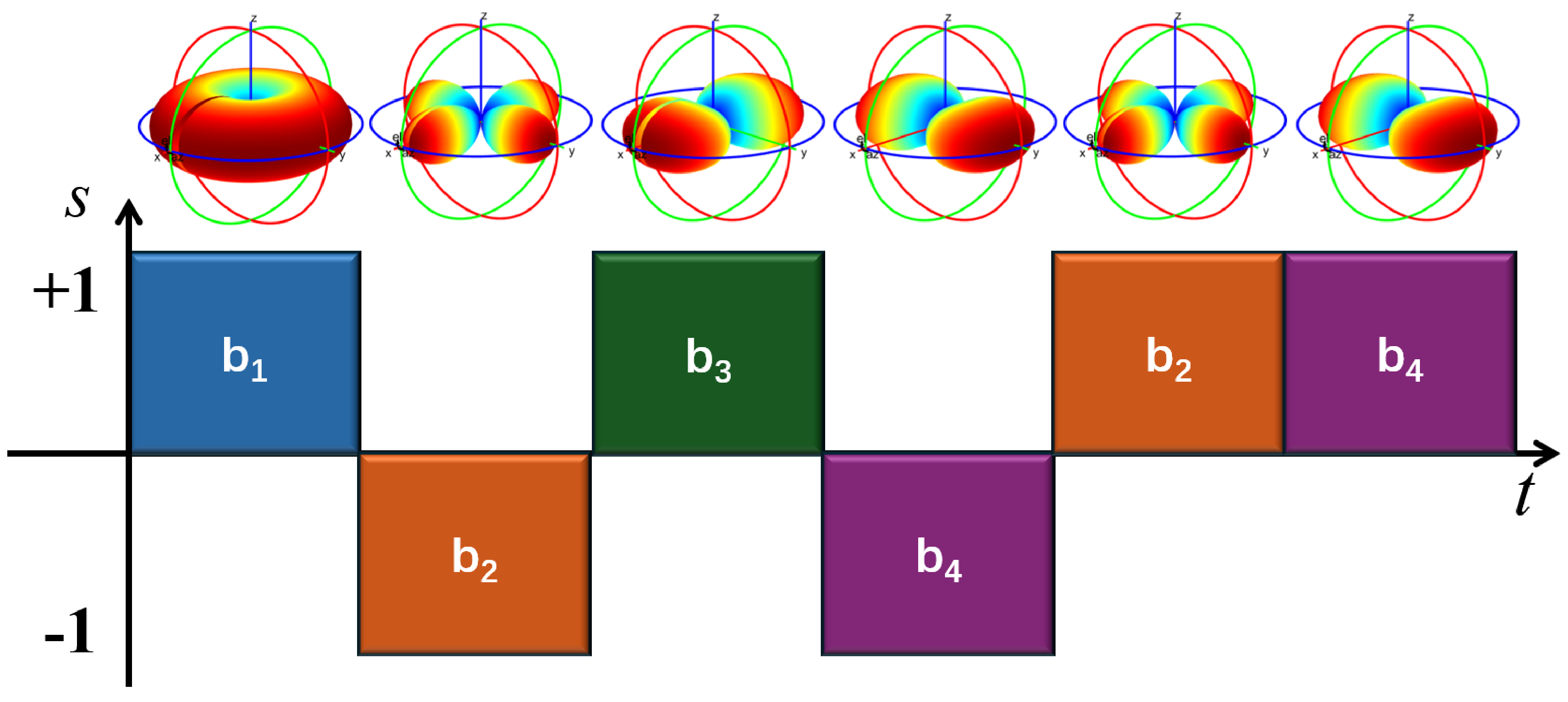}}
\par\end{centering}
\begin{centering}
(b)
\par\end{centering}
\caption{Illustration of (a) MIMO transmitter with pixel antennas and (b) transmission
scheme with the fast antenna coding.\label{fig:MIMO}}
\end{figure}
We can decompose $\mathbf{E}_{\mathrm{oc}}$ by singular value decomposition
as $\mathbf{E}_{\mathrm{oc}}=\mathbf{U}\mathbf{\Sigma}\mathbf{V}^{H}$
where each column in $\mathbf{U}$ provides an orthogonal basis radiation
pattern. $\mathbf{\Sigma}=\mathrm{diag}\left(\sigma_{1},\sigma_{2},...,\sigma_{R}\right)\in\mathbb{R}^{R\times R}$
is the singular value matrix where $R=\mathrm{rank}\left(\mathbf{E}_{\mathrm{oc}}\right)$
is the rank of matrix $\mathbf{E}_{\mathrm{oc}}$ and refers to the
effective aerial DoF (EADoF). As a result, $\mathbf{e}\left(\mathbf{b}\right)$
is given by
\begin{align}
\mathbf{e}\left(\mathbf{b}\right) & =\mathbf{U}\mathbf{\Sigma}\mathbf{V}^{H}\mathbf{i}\left(\mathbf{b}\right)=\mathbf{U}\mathbf{w}\left(\mathbf{b}\right),\label{eq:pattern coder}
\end{align}
where $\mathbf{w}\left(\mathbf{b}\right)=\mathbf{\Sigma}\mathbf{V}^{H}\mathbf{i}\left(\mathbf{b}\right)\in\mathbb{C}^{R\times1}$
is referred to as the \textit{pattern coder} which is composed of
weighting coefficients to linearly code the orthogonal basis radiation
patterns in $\mathbf{U}$ and synthesize radiation pattern of the
pixel antenna.

\section{Fast Antenna Coding for MIMO Systems}

Considering a MIMO system employing $N$ pixel antennas at the transmitter,
as shown in Fig. \ref{fig:MIMO}(a), each pixel antenna is connected
to an RF chain. Radiation patterns of $N$ pixel antennas are collected
into a matrix $\mathbf{E}\left(\mathbf{B}\right)=\left[\mathbf{e}_{1}\left(\mathbf{b}_{1}\right),\mathbf{e}_{2}\left(\mathbf{b}_{2}\right),...,\mathbf{e}_{N}\left(\mathbf{b}_{N}\right)\right]\in\mathbb{C}^{2K\times N}$
with $\left\Vert \mathbf{e}_{n}\left(\mathbf{b}_{n}\right)\right\Vert ^{2}=1$
where $\mathbf{B}=\left[\mathbf{b}_{1},\mathbf{b}_{2},...,\mathbf{b}_{N}\right]\in\mathbb{R}^{Q\times N}$
is the matrix collecting antenna coders for $N$ pixel antennas. At
the receiver side, $M$ receive antennas have fixed configurations
and their radiation patterns $\mathbf{f}_{m}$ with $\left\Vert \mathbf{f}_{m}\right\Vert ^{2}=1$,
$\forall m$, are collected into a matrix $\mathbf{F}=\left[\mathbf{f}_{1},\mathbf{f}_{2},\ldots,\mathbf{f}_{M}\right]\in\mathbb{C}^{2K\times M}$.
Using the beamspace channel representation \cite{sayeed2002deconstructing},
we can include the effect of radiation patterns into the propagation
environment as
\begin{equation}
\mathbf{H}\left(\mathbf{B}\right)=\mathbf{F}^{T}\mathbf{H}_{\mathrm{V}}\mathrm{\mathbf{E}}\left(\mathbf{B}\right).\label{eq:H Current}
\end{equation}
where $\mathbf{H}_{\mathrm{V}}\in\mathbb{C}^{2K\times2K}$ is the
virtual channel matrix whose entries refer to the channel gain from
each angle of departure to each angle of arrival among $K$ samples
for two polarization components. Particularly in a rich scattering
environment, we assume entries in $\mathbf{H}_{\mathrm{V}}$ follow
the i.i.d. complex Gaussian distribution $\mathcal{CN}\left(0,1\right)$
.

We collect orthogonal basis radiation patterns of $N$ pixel antennas
to form the beamspace $\mathrm{\mathbf{U}_{BS}}=\left[\mathbf{U}_{1},\mathbf{U}_{2},...,\mathbf{U}_{N}\right]\in\mathbb{C}^{2K\times NR}$
and collect all pattern coders $\mathbf{w}_{n}\left(\mathbf{b}_{n}\right)=\mathbf{\Sigma}_{n}\mathbf{V}^{H}_{n}\mathbf{i}_{n}\left(\mathbf{b}_{n}\right)$
into a block diagonal matrix $\mathbf{W}\left(\mathbf{\mathbf{B}}\right)=\mathrm{blkdiag}\left(\mathbf{w}_{1}\left(\mathbf{b}_{1}\right),\mathbf{w}_{2}\left(\mathbf{b}_{2}\right),...,\mathbf{w}_{N}\left(\mathbf{b}_{N}\right)\right)\in\mathbb{\mathbb{C}}^{NR\times N}$.
Then the system model can be expressed as
\begin{align}
\mathbf{y} & =\mathbf{H}\left(\mathbf{B}\right)\mathbf{s}+\mathbf{n}=\mathbf{F}^{T}\mathbf{H}_{\mathrm{V}}\mathbf{U}_{\mathrm{BS}}\mathbf{W}\left(\mathbf{B}\right)\mathbf{s}+\mathbf{n}\nonumber \\
 & =\mathbf{F}^{T}\mathbf{H}_{\mathrm{V}}\mathbf{U}_{\mathrm{BS}}\mathbf{x}\left(\mathbf{B},\mathbf{s}\right)+\mathbf{n}=\mathbf{H}_{\mathrm{BS}}\mathbf{x}\left(\mathbf{B},\mathbf{s}\right)+\mathbf{n},\label{eq:BS system}
\end{align}
where $\mathbf{s}=\left[s_{1},s_{2},...,s_{N}\right]\in\mathbb{C}^{N\times1}$
collects the baseband symbols, $s_{n}$ for $n=1,2,...,N$, output
by the $N$ RF chains into a vector, $\mathbf{x}\left(\mathbf{B},\mathbf{s}\right)=\mathbf{W}\left(\mathbf{B}\right)\mathbf{s}\in\mathbb{C}^{NR\times1}$
denotes the equivalent beamspace transmit signal and $\mathbf{H}_{\mathrm{BS}}=\mathbf{F}^{H}\mathbf{H}_{\mathrm{V}}\mathbf{U}_{\mathrm{BS}}\in\mathbb{C}^{M\times NR}$
denotes the equivalent beamspace channel matrix. We note here that
entries in $\mathbf{H}_{\mathrm{BS}}$ follow the i.i.d. complex Gaussian
distribution $\mathcal{CN}\left(0,1\right)$ since $\mathbf{F}^{H}\mathbf{F}=\mathbf{I}_{M}$
and $\mathbf{U}^{H}_{\mathrm{BS}}\mathbf{U}_{\mathrm{BS}}=\mathbf{I}_{NR}$.

We can observe from \eqref{eq:BS system} that $\mathbf{B}$ and $\mathbf{s}$
are mutually independent in $\mathbf{x}\left(\mathbf{B},\mathbf{s}\right)$
so that $N$ pixel antennas can simultaneously transmit the baseband
modulated symbol $\mathbf{s}$ and perform antenna coding by $\mathbf{B}$
to transmit information bits via selecting radiation pattern, as shown
in Fig. \ref{fig:MIMO}(b). In this case, the switching interval of
antenna coders is at the same level of the baseband symbol period
and this is referred to as the \textit{fast antenna coding}.

To analyze SE of MIMO system with the fast antenna coding, we consider
a codebook composed of $P$ different antenna coders $\mathcal{B}=\left\{ \mathbf{b}_{\mathrm{C},1},\mathbf{b}_{\mathrm{C},2},...,\mathbf{b}_{\mathrm{C},P}\right\} $
so that the antenna coder of each pixel antenna, $\mathbf{b}_{n}$,
$\forall n$, can be selected from this codebook. The $P$ indices
of antenna coders can be mapped into information bits which are then
modulated in the beamspace and transmitted by the corresponding radiation
patterns. Now we can write the mutual information (MI) of the MIMO
system with the fast antenna coding \eqref{eq:BS system}, denoted
as $\mathrm{I}\left(\mathbf{x}\left(\mathbf{B},\mathbf{s}\right);\mathbf{y}|\mathbf{H}_{\mathrm{BS}}\right)$,
which is given by \cite{6877673}
\begin{align}
\mathrm{I}\left(\mathbf{x}\left(\mathbf{B},\mathbf{s}\right);\mathbf{y}|\mathbf{H}_{\mathrm{BS}}\right) & =\mathrm{I}\left(\mathbf{s};\mathbf{y}|\mathbf{W}\left(\mathbf{B}\right),\mathbf{H}_{\mathrm{BS}}\right)\nonumber \\
 & \:\quad+\mathrm{I}\left(\mathbf{W}\left(\mathbf{B}\right);\mathbf{y}|\mathbf{H}_{\mathrm{BS}}\right).\label{eq:mutual information}
\end{align}
The first term in \eqref{eq:mutual information} is the MI from transmit
baseband symbol $\mathbf{s}$, which is equivalent to the MI of a
conventional MIMO as
\begin{align}
 & \mathrm{I}\left(\mathbf{s};\mathbf{y}|\mathbf{W}\left(\mathbf{B}\right),\mathbf{H}_{\mathrm{BS}}\right)=\mathrm{I}\left(\mathbf{s};\mathbf{y}|\mathbf{H}_{\mathrm{BS}}\mathbf{W}\left(\mathbf{B}\right)\right)\nonumber \\
 & =\mathrm{\log}_{2}\left|\mathbf{I}_{M}+\frac{\mathbf{H}_{\mathrm{BS}}\mathbf{W}\left(\mathbf{B}\right)\mathrm{E}\left[\mathbf{s}\mathbf{s}^{H}\right]\mathbf{W}^{H}\left(\mathbf{B}\right)\mathbf{H}^{H}_{\mathrm{BS}}}{N_{0}}\right|.\label{eq:MIMO MI}
\end{align}
The second term in \eqref{eq:mutual information} is the MI of fast
antenna coding. When the antenna coders are randomly selected with
equal probability, the second term is given by
\begin{align}
 & \mathrm{I}\left(\mathbf{W}\left(\mathbf{B}\right);\mathbf{y}|\mathbf{H}_{\mathrm{BS}}\right)\nonumber \\
 & =\mathrm{H}\left(\mathbf{W}\left(\mathbf{B}\right)|\mathbf{H}_{\mathrm{BS}}\right)-\mathrm{H}\left(\mathbf{W}\left(\mathbf{B}\right)|\mathbf{H}_{\mathrm{BS}},\mathbf{y}\right)\nonumber \\
 & =N\mathrm{\log}_{2}P-\mathrm{E}_{\mathbf{W}\left(\mathbf{B}\right),\mathbf{H}_{\mathrm{BS}},\mathbf{y}}\left[\mathrm{\mathrm{\log}_{2}}\textrm{prob}\left(\mathbf{W}\left(\mathbf{B}\right)|\mathbf{H}_{\mathrm{BS}},\mathbf{y}\right)^{-1}\right]\nonumber \\
 & =N\mathrm{\log}_{2}P-\frac{\mathrm{E}_{\mathbf{s}}\left[\mathrm{exp}\left(-\frac{\left\Vert \mathbf{y}-\mathbf{H}_{\mathrm{BS}}\mathbf{W}\left(\mathbf{B}\right)\mathbf{s}\right\Vert ^{2}}{N_{0}}\right)\right]}{\sum^{P^{N}}_{i=1}\mathrm{E}_{\mathbf{s}}\left[\mathrm{exp}\left(-\frac{\left\Vert \mathbf{y}-\mathbf{H}_{\mathrm{BS}}\mathbf{W}\left(\mathbf{B}_{i}\right)\mathbf{s}\right\Vert ^{2}}{N_{0}}\right)\right]}.\label{eq:AC MI}
\end{align}
Therefore, by using the fast antenna coding technique, SE of MIMO
systems can be enhanced by \eqref{eq:AC MI}, which approaches $N\mathrm{\log}_{2}P$
when the signal-to-noise ratio (SNR) is large.

The BER of MIMO with fast antenna coding is given by
\begin{align}
\mathrm{BER}= & \frac{\sum_{\left(i^{\prime},j^{\prime}\right)}p\left(\mathbf{x}\left(\mathbf{B}_{i},\mathbf{s}_{j}\right)\rightarrow\mathbf{x}\left(\mathbf{B}_{i^{\prime}},\mathbf{s}_{j^{\prime}}\right)|\mathbf{H}_{\mathrm{BS}}\right)}{\left(PM\right)^{N}N\mathrm{log}_{2}\left(PM\right)}\cdot\nonumber \\
 & d\left(\mathbf{x}\left(\mathbf{B}_{i},\mathbf{s}_{j}\right)-\mathbf{x}\left(\mathbf{B}_{i^{\prime}},\mathbf{s}_{j^{\prime}}\right)\right)\label{eq:BER}
\end{align}
where $M$-ary digital modulation is performed on the baseband symbols
in $\mathbf{s}$, $p\left(\mathbf{x}\left(\mathbf{B}_{i},\mathbf{s}_{j}\right)\rightarrow\mathbf{x}\left(\mathbf{B}_{i^{\prime}},\mathbf{s}_{j^{\prime}}\right)|\mathbf{H}_{\mathrm{BS}}\right)=Q\left(\sqrt{\frac{\left\Vert \mathbf{H}_{\mathrm{BS}}\left(\mathbf{x}\left(\mathbf{B}_{i},\mathbf{s}_{j}\right)-\mathbf{x}\left(\mathbf{B}_{i^{\prime}},\mathbf{s}_{j^{\prime}}\right)\right)\right\Vert ^{2}}{2\sigma^{2}}}\right)$
is the pairwise error probability (PEP) between two equivalent beamspace
transmit signals $\mathbf{x}\left(\mathbf{B}_{i},\mathbf{s}_{j}\right)$
and $\mathbf{x}\left(\mathbf{B}_{i^{\prime}},\mathbf{s}_{j^{\prime}}\right)$
for $\left(i,j\right)\neq\left(i^{\prime},j^{\prime}\right)$, $d\left(\left(\mathbf{B}_{i},\mathbf{s}_{j}\right),\mathbf{x}\left(\mathbf{B}_{i^{\prime}},\mathbf{s}_{j^{\prime}}\right)\right)$
is the distance between the information bits carried by $\mathbf{x}\left(\mathbf{B}_{i},\mathbf{s}_{j}\right)$
and $\mathbf{x}\left(\mathbf{B}_{i^{\prime}},\mathbf{s}_{j^{\prime}}\right)$.

\section{Antenna Coder Optimization}

In this section, we provide an efficient optimization method to obtain
the optimal antenna coders for the fast antenna coding. We can optimize
the $P$ antenna coders in the codebook $\mathcal{B}=\left\{ \mathbf{b}_{\mathrm{C},1},\mathbf{b}_{\mathrm{C},2},...,\mathbf{b}_{\mathrm{C},P}\right\} $
so that each transmit pixel antenna can select among $P$ radiation
patterns to modulate information bits. Specifically, given the ideal
pattern coders $\mathcal{W}=\left\{ \mathbf{w}_{\mathrm{C},1},\mathbf{w}_{\mathrm{C},2},...,\mathbf{w}_{\mathrm{C},P}\right\} $
that modulate the information bits onto the radiation patterns in
the beamspace, our aim is to find the actual pattern coders $\mathbf{w}\left(\mathbf{b}_{\mathrm{C},p}\right)$
that is most similar to $\mathbf{w}_{\mathrm{C},p}$, $\forall p$.
To that end, we consider the correlation coefficient between $\mathbf{w}\left(\mathbf{b}_{\mathrm{C},p}\right)$
and $\mathbf{w}_{\mathrm{C},p}$ defined as
\begin{equation}
\rho_{p}\left(\mathbf{b}_{\mathrm{C},p}\right)=\mathbf{w}^{H}\left(\mathbf{b}_{\mathrm{C},p}\right)\mathbf{w}_{\mathrm{C},p}.\label{eq:Corr Coef}
\end{equation}
It should be noted that both $\mathbf{w}\left(\mathbf{b}_{\mathrm{C},p}\right)$
and $\mathbf{w}_{\mathrm{C},p}$ have unity norm so that the correlation
coefficient satisfies $0\leqslant\left|\rho_{p}\left(\mathbf{b}_{\mathrm{C},p}\right)\right|\leqslant1$.
When $\left|\rho_{p}\left(\mathbf{b}_{\mathrm{C},p}\right)\right|$
is close to unity, $\mathbf{w}\left(\mathbf{b}_{\mathrm{C},p}\right)$
is almost the same as $\mathbf{w}_{\mathrm{C},p}$. In this case,
the phase differences between $\mathbf{w}\left(\mathbf{b}_{\mathrm{C},p}\right)$
and $\mathbf{w}_{\mathrm{C},p}$ can be compensated by the output
signal on the RF chain. Leveraging the correlation coefficient, finding
the optimal antenna coder can be performed by solving the optimization
problem
\begin{align}
\underset{\mathbf{b}_{\mathrm{C},p},\forall p}{\mathrm{max}}\:\: & \frac{1}{P}\stackrel[p=1]{P}{\sum}\left|\rho_{p}\left(\mathbf{b}_{\mathrm{C},p}\right)\right|^{2}\label{eq:problem}\\
\mathrm{s.t.}\:\:\:\:\: & \mathbf{b}_{\mathrm{C},p}\in\left\{ 0,1\right\} ^{Q},\forall p,\label{eq:subj}
\end{align}
which maximizes the mean correlations between all pairs of actual
pattern coders and ideal pattern coders. To solve the binary optimization
problem \eqref{eq:problem}-\eqref{eq:subj}, we can use the genetic
algorithm (GA) \cite{Zhang2022} to obtain the optimal codebook $\mathcal{B}^{\star}=\left\{ \mathbf{b}^{\star}_{\mathrm{C},1},\mathbf{b}^{\star}_{\mathrm{C},2},...,\mathbf{b}^{\star}_{\mathrm{C},P}\right\} $.

\section{Numerical Simulations}

\begin{figure}[t]
\begin{centering}
\includegraphics[width=6cm]{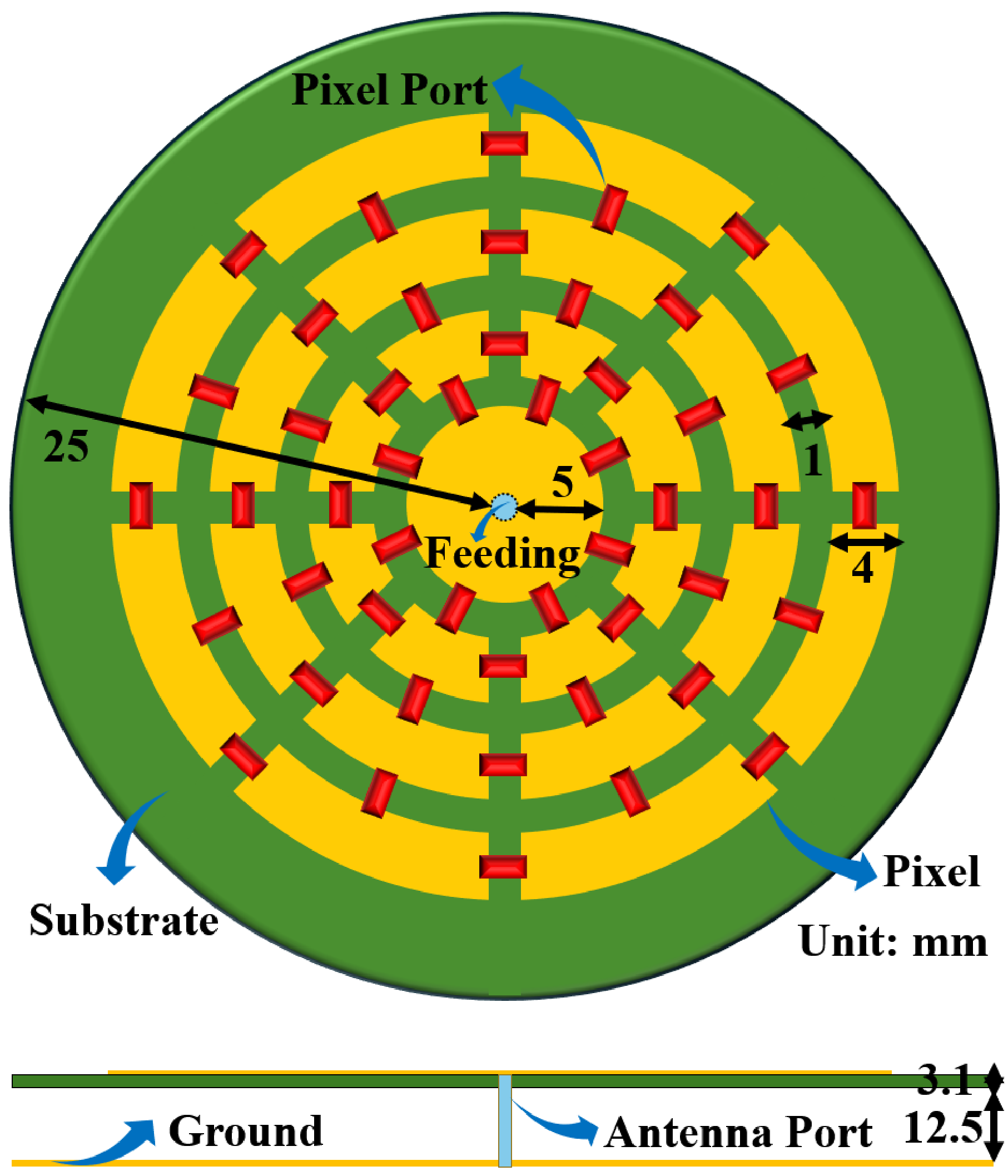}, 
\par\end{centering}
\caption{Plan and elevation views of the pixel antenna design (unit: mm). \label{fig:Design}}
\end{figure}
In this section, we evaluate the performance of MIMO systems with
the fast antenna coding based on pixel antennas.

\subsection{Pixel Antenna Design and Radiation Patterns}

\begin{figure}[t]
\begin{centering}
\textsf{\includegraphics[width=8.8cm]{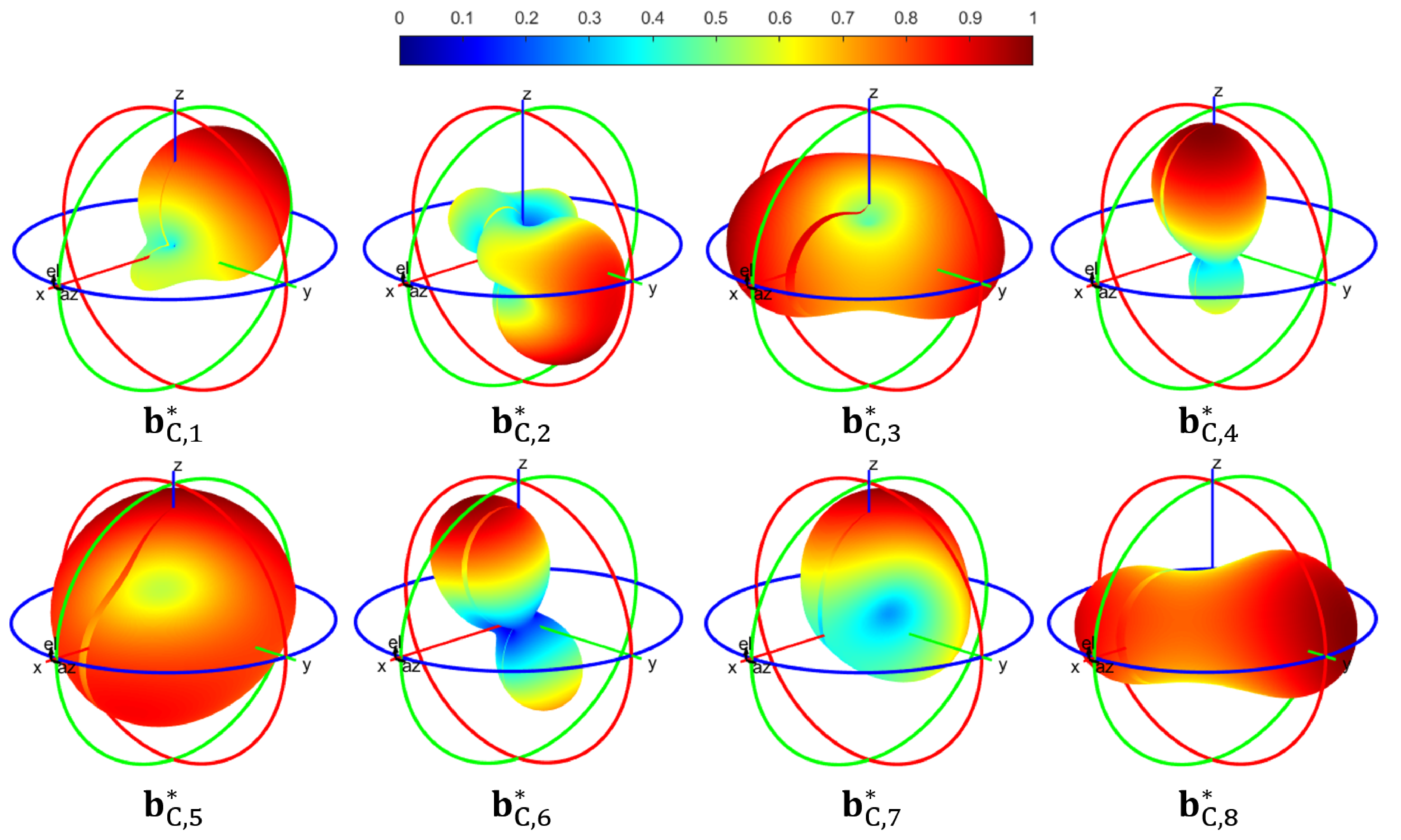}}
\par\end{centering}
\caption{Radiation patterns of the pixel antenna by using $P=8$ optimized
antenna coders.\label{fig:Pattern}}
\end{figure}

Inspired by \cite{Zhang2022}, we design the pixel antenna in a circular
form, as shown in Fig. \ref{fig:Design}, which has $Q=48$ switches.
The pixels and substrate are made of copper and Rogers 4003C, respectively.
The full 3-dimensional radiation pattern with uniform power angular
spectrum is considered. The proposed design has four dominant orthogonal
basis radiation patterns which contribute over 95\% of the total radiated
power. Therefore in this work, the EADoF of the proposed pixel antenna
is $R=4$. Then we construct the codebook with $P=8$ ideal pattern
coders as shown in Table \ref{tab:Pattern Coder}, where different
orthogonal basis radiation patterns can be excited by using $\mathbf{w}_{\mathrm{C},1}$
to $\mathbf{w}_{\mathrm{C},4}$ while $\mathbf{w}_{\mathrm{C},5}$
to $\mathbf{w}_{\mathrm{C},8}$ combine two orthogonal basis radiation
patterns with equal weight. Then by using the optimization approach
in Section IV, we obtain eight antenna coders $\mathcal{B}^{\star}=\left\{ \mathbf{b}^{\star}_{\mathrm{C},1},\mathbf{b}^{\star}_{\mathrm{C},2},...,\mathbf{b}^{\star}_{\mathrm{C},8}\right\} $
for the proposed pixel antenna to excite the radiation patterns as
shown in Fig. \ref{fig:Pattern}. We can modulate additional information
bits on these radiation patterns to enhance SE as illustrated in Fig.
\ref{fig:MIMO}.

\begin{table}[t]
\caption{$P=8$ ideal pattern coders for fast antenna coding \label{tab:Pattern Coder}}

\centering{}%
\begin{tabular}{c|c}
\hline 
$\mathbf{w}_{\mathrm{C},1}=\left[1,0,0,0\right]$ & $\mathbf{w}_{\mathrm{C},5}=\frac{1}{\sqrt{2}}\left[1,0,1,0\right]$\tabularnewline
\hline 
$\mathbf{w}_{\mathrm{C},2}=\left[0,1,0,0\right]$ & $\mathbf{w}_{\mathrm{C},6}=\frac{1}{\sqrt{2}}\left[0,1,0,1\right]$\tabularnewline
\hline 
$\mathbf{w}_{\mathrm{C},3}=\left[0,0,1,0\right]$ & $\mathbf{w}_{\mathrm{C},7}=\frac{1}{\sqrt{2}}\left[1,0,0,1\right]$\tabularnewline
\hline 
$\mathbf{w}_{\mathrm{C},4}=\left[0,0,0,1\right]$ & $\mathbf{w}_{\mathrm{C},8}=\frac{1}{\sqrt{2}}\left[0,1,1,0\right]$\tabularnewline
\hline 
\end{tabular}
\end{table}

\subsection{SE and EE Performance}

\begin{figure}[t]
\begin{centering}
\textsf{\includegraphics[width=8.5cm]{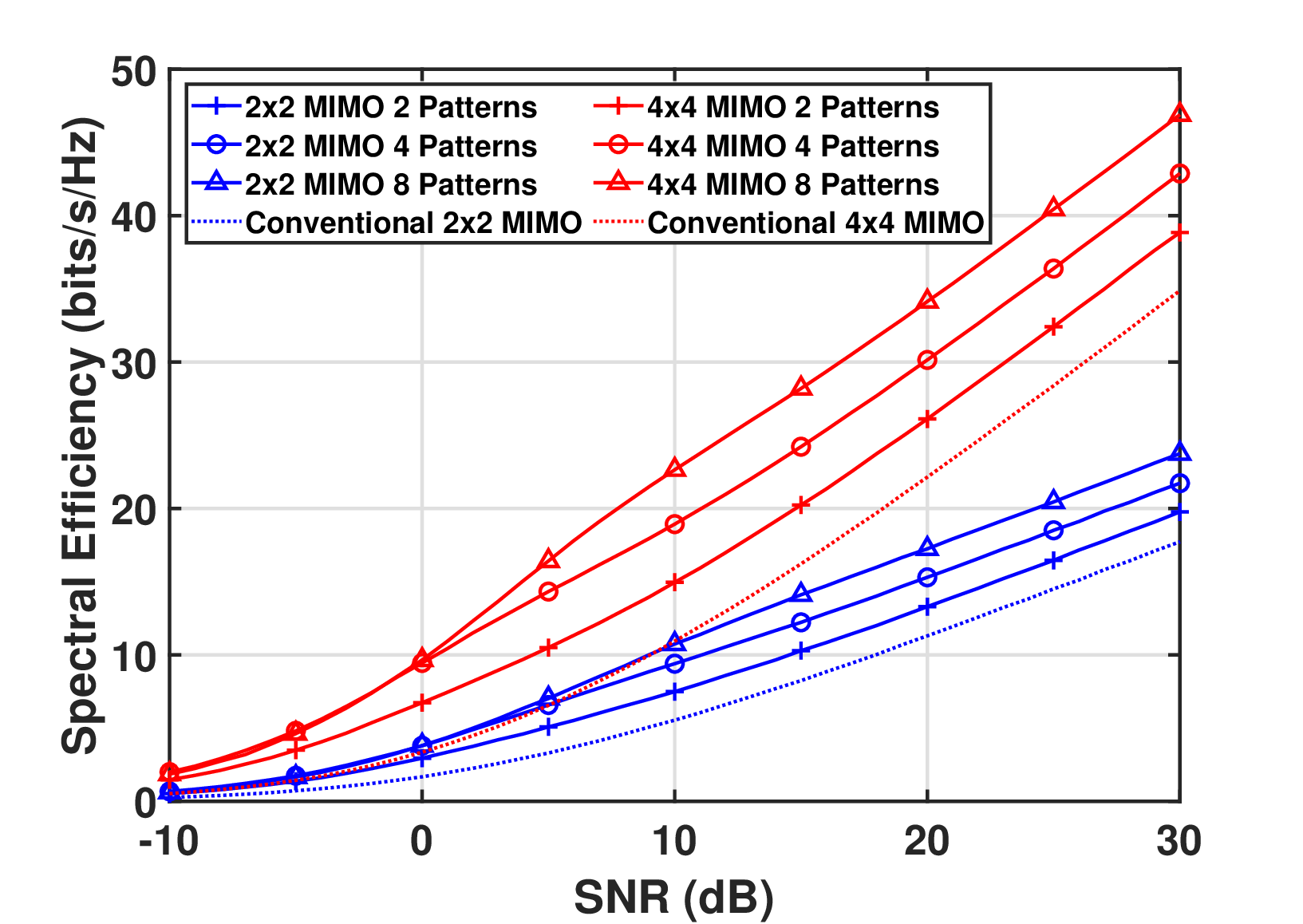}}
\par\end{centering}
\begin{centering}
(a)
\par\end{centering}
\begin{centering}
\textsf{\includegraphics[width=8.5cm]{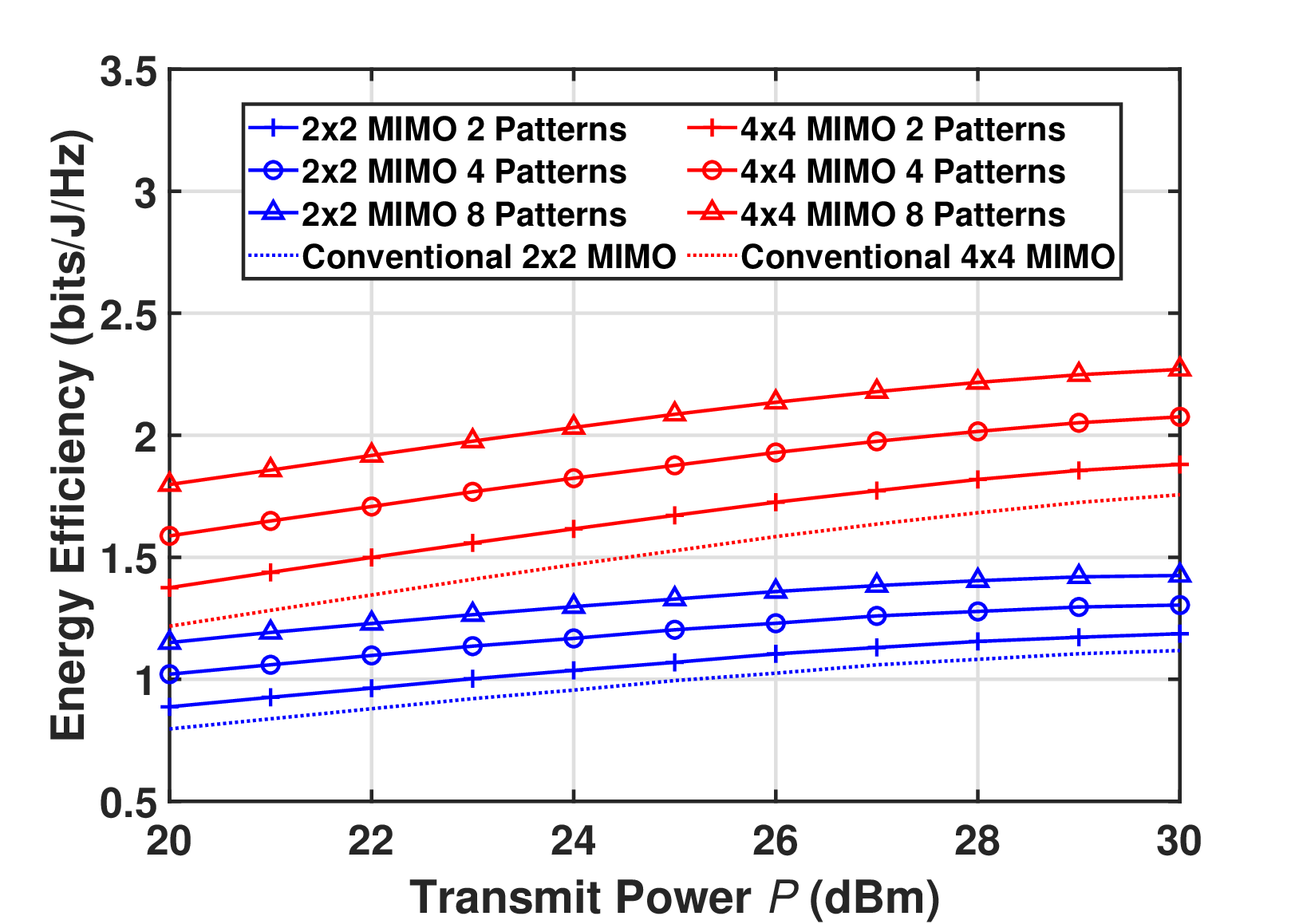}}
\par\end{centering}
\begin{centering}
(b)
\par\end{centering}
\caption{(a) SE and (b) EE of MIMO systems with the fast antenna coding.\label{fig:SE}}
\end{figure}

To verify the performance of the fast antenna coding, we utilize radiation
patterns in Fig. \ref{fig:Pattern} in MIMO systems which employ pixel
antennas at the transmitters and conventional fixed antennas at the
receivers. The simulated SE of $2\times2$ and $4\times4$ MIMO systems
with $2$, $4$ and $8$ radiation patterns are shown in Fig. \ref{fig:SE}
(a). Particularly, for the 2 and 4-pattern cases, we use the optimized
antenna coders $\left\{ \mathbf{b}^{\star}_{\mathrm{C},1},\mathbf{b}^{\star}_{\mathrm{C},2}\right\} $
and $\left\{ \mathbf{b}^{\star}_{\mathrm{C},1},\mathbf{b}^{\star}_{\mathrm{C},2},\mathbf{b}^{\star}_{\mathrm{C},3},\mathbf{b}^{\star}_{\mathrm{C},4}\right\} $,
respectively, to modulate additional information bits on the corresponding
radiation patterns. From Fig. \ref{fig:SE}, we can observe that SE
of MIMO systems with the fast antenna coding can be significantly
enhanced when compared with conventional MIMO systems with fixed antenna
configurations. Particularly, SE is enhanced by 8 and 12 bits/s/Hz
(equivalent to 22\% and 33\%) in $4\times4$ MIMO system when 4 and
$8$ antenna coders are utilized, respectively. In addition, by increasing
the number of antenna coders, SE of MIMO systems with the fast antenna
coding can be improved when SNR is higher than 0 dB, since more information
bits can be modulated on the radiation patterns of pixel antennas.
However, when SNR is lower than 0 dB, the SE improvement from the
4-pattern case to 8-pattern case is minor since the pattern coders
are mutually correlated rather than orthogonal. Further, SE of $4\times4$
MIMO system is twice that of $2\times2$ MIMO system when using the
fast antenna coding. This is because the fast antenna coding introduces
extra DoF in the beamspace domain while not affecting the original
spatial multiplexing of MIMO systems. These results show the effectiveness
of the fast antenna coding based on pixel antennas in enhancing SE
of MIMO systems.

We also simulate EE, defined as the ratio of SE to power consumption,
of MIMO systems with the fast antenna coding as shown in Fig. \ref{fig:SE}
(b). We can observe that EE of MIMO systems can be significantly improved,
by up to 33\%, by using the fast antenna coding, compared with conventional
MIMO systems using fixed antenna configurations, owing to the dominant
SE enhancement although power consumption of RF switches is taken
into consideration.

\subsection{BER Performance}

\begin{figure}[t]
\begin{centering}
\textsf{\includegraphics[width=8.5cm]{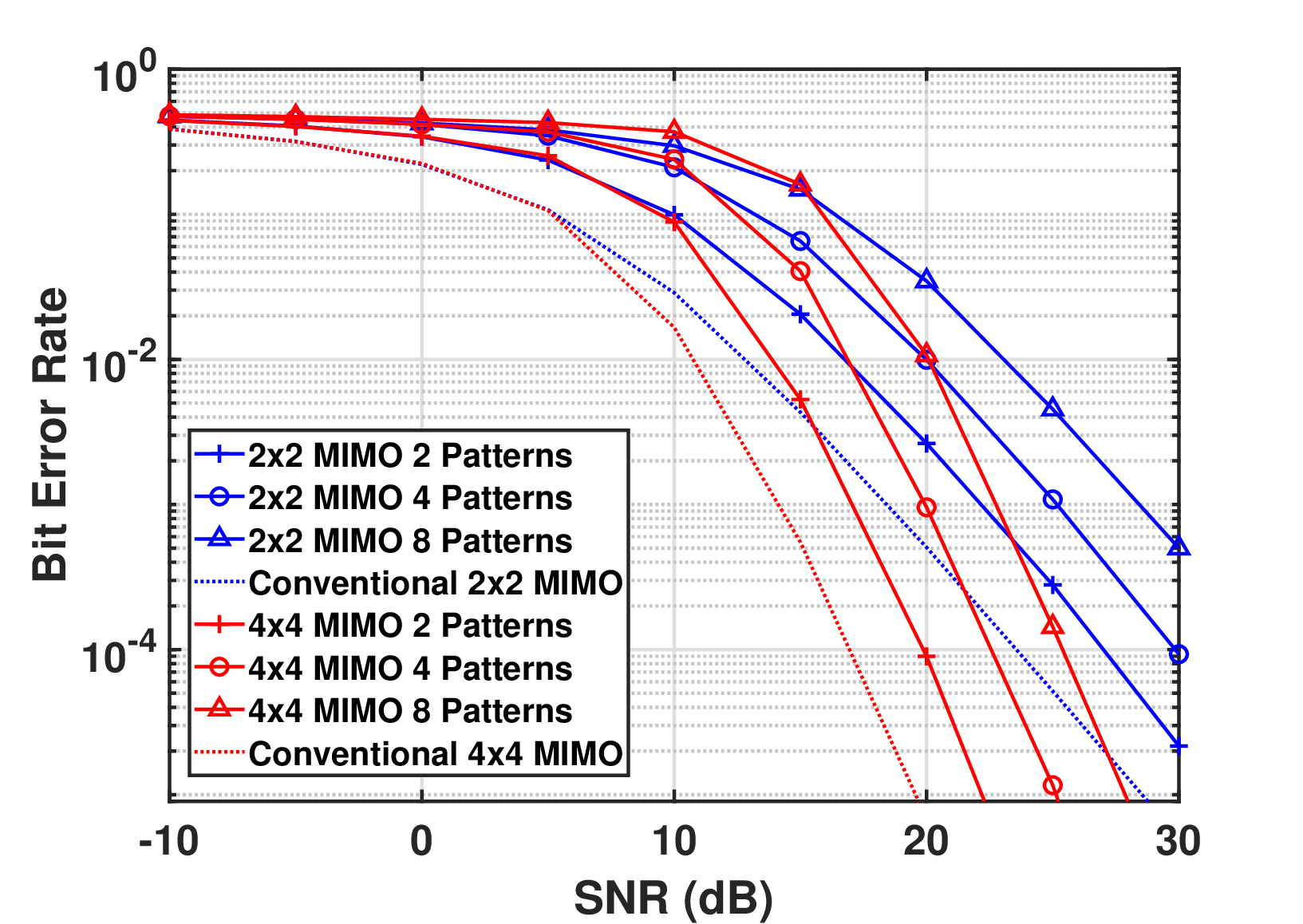}}
\par\end{centering}
\caption{BER performance of MIMO systems with the fast antenna coding.\label{fig:BER}}
\end{figure}
Finally, we simulate BER performance of MIMO systems with the fast
antenna coding where we use quadrature phase shift keying (QPSK) modulation
for baseband symbols and utilize the maximum likelihood detector at
the receiver side. As shown in Fig. \ref{fig:BER}, we can observe
that for MIMO with the fast antenna coding, using more radiation patterns
to transmit additional information bits will increase BER simultaneously.
Specifically in the $4\times4$ MIMO system, the SNR gap, when achieving
the same BER performance, between MIMO systems with the fast antenna
coding and the conventional MIMO system is around 5 dB and 9 dB for
4-pattern and 8-pattern cases respectively. Therefore, there is a
tradeoff between SE and BER when performing the fast antenna coding
in the MIMO system. That is, the fast antenna coding with more antenna
coders can be performed when the received SNR is sufficiently high
to maintain an acceptable BER. Conversely, the number of radiation
patterns should be limited when the received SNR is low.

\section{Conclusions}

In this work, we investigate the fast antenna coding technique based
on pixel antennas for enhancing the SE of MIMO systems. The proposed
technique introduces additional DoF in the beamspace domain for transmitting
information bits, thus breaking through the SE bound of conventional
MIMO systems. This is achieved by controlling the antenna coders to
modulate information on the radiation patterns of pixel antennas.
Specifically, the concept of the fast antenna coding based on pixel
antennas is described. Then the SE and BER of MIMO systems with the
fast antenna coding are provided, showing the SE enhancement by leveraging
fast antenna coding. By utilizing the GA approach, the antenna coders
are optimized to obtain the desired pattern coders for fast antenna
coding. Finally, we demonstrate the fast antenna coding by using antenna
coders to excite various radiation patterns of pixel antennas in a
MIMO system. It is shown that in the $4\times4$ MIMO system, SE is
enhanced by 12 bits/s/Hz and the EE is improved by 33\% when compared
with conventional MIMO, showing the effectiveness of the fast antenna
coding.

\section*{Acknowledgment}

This work was funded by China Mobile Research Institute, by the Science
and Technology Development Fund, Macau SAR (File/Project no. 001/2024/SKL
and 0002/2025/EQP), by University of Macau (File no. SRG2025-00060-IOTSC),
and by University of Macau Development Foundation (UMDF) (File no.
UMDF-TISF-I/2026/025/IOTSC).

\bibliographystyle{IEEEtran}

\end{document}